\documentclass[10pt,twoside]{hsqcd}
\usepackage{epsf,amsmath}
\usepackage{graphicx}
\usepackage{color}

\begin{document}

\title{ANGULAR DISTRIBUTIONS OF DRELL-YAN LEPTONS IN THE PARTON REGGEIZATION APPROACH
\thanks{The
work  was supported by the Ministry for Science and Education of the
Russian Federation under Contract No.~14.740.11.0894.}}

\author{\underline{M.~A.~NEFEDOV} and V.~A.~SALEEV\\ \\
Samara State University \\
Samara 443011, Russia\\
E-mail: saleev@samsu.ru, nefedovma@gmail.com}

\maketitle

\begin{abstract}
We study angular distributions of Drell-Yan leptons in the
proton-proton  collisions invoking the hypothesis of quark
Reggeization in $t$-channel exchanges at high energy.
\end{abstract}



\markboth{\large \sl \underline{M.~A.~NEFEDOV} \& V.~A.~SALEEV
\hspace*{2cm} HSQCD 2012} {\large \sl \hspace*{1cm} TEMPLATE FOR THE
HSQCD 20012 PROCEEDINGS}

\section{Introduction}
\label{sec:one}

The inclusive production of two leptons with opposite electric
charge $l^+l^-$ and invariant mass $Q$ larger than a few GeV, which
is usually specify as Drell-Yan pair production \cite{DrellYan}, in
hadronic collisions is considered as very important process of
high-energy hadronic physics. The experimental study of Drell-Yan
pair production includes measurements of massive lepton pair
distributions in rapidity ($y$), invariant mass ($Q$), transverse
momentum ($q_T=|{\vec q}_T|$), and angular distributions of leptons
in the virtual photon rest frame. The last ones are usually
presented in terms of so-called angular coefficients, which is being
a subject of study in this paper. The relevant data for
proton-proton collisions have been obtained at the Fermilab Tevatron
by NuSea Collaboration \cite{NuSea} at the $\sqrt{S}=39$ GeV.
Theoretical study of Drell-Yan pair production is based on the
Parton Model (PM) and  perturbative quantum chromodynamics (QCD) in
leading-order (LO) and next-to-leading-order (NLO) approximations
\cite{Review}, as well as on the soft initial-state gluon
resummation procedure in all orders in $\alpha_s$ \cite{Berger2}
(see, also, references therein). The attempts of description of
Drell-Yan pair production   were undertaken  also in the uncollinear
factorization scheme, namely $k_T-$factorization approach, which
takes into account off-shell properties of $t-$channel exchange
partons and unintegrated parton PDFs \cite{KT1,KT2}. In this paper
we study Drell-Yan pair production in the framework of the Parton
Reggeization Approach (PRA)\cite{Lipatov95,BFKL}, invoking the
hypothesis of quark Reggeization in $t$-channel exchanges at high
energy \cite{FadinSherman}. The quark Reggeization hypothesis has
been used successfully for description of different spectra of
prompt photons at the Fermilab Tevatron and CERN LHC
\cite{tevatronY,lhcY}, electron deep inelastic scattering and prompt
photon production cross sections at the DESY HERA
\cite{tevatronY,heraY}.

\section{Drell-Yan pair production in QCD}
\label{sec:two}

In the experiments under consideration, massive lepton pairs are
produced with substantial transverse momentum $\vec{q}_T$. To
describe nonzero transverse momentum Drell-Yan pair production we
need to consider in the collinear PM the NLO partonic subprocesses
$2\to 2$: $q+\bar q\to g + \gamma^\star\to g+l^++ l^-$, $q+g\to
q+\gamma^\star\to q+l^+ +l^-$. The Drell-Yan pair production in
collisions of  hadrons have been study carefully in the NLO QCD and
the collinear PM, excluding the region of small $q_T$ and $Q$
\cite{Review}. In the $k_T-$factorization approach \cite{KT1,KT2}
the processes with the off-shell initial partons $(q^*, g^*)$ having
nonzero transverse momenta are considered as the source of the LO
and NLO contributions: $ q^*+\bar q^*\to \gamma^\star\to l^++ l^-$,
$q^*+\bar q^*\to g + \gamma^\star\to g+l^++ l^-$, $q^*+g^*\to
q+\gamma^\star\to q+l^+ +l^-$. However, the conception of off-shell
quarks {is not defined correctly} in the $k_T-$factorization
approach because it {breaks} the gauge invariance of relevant
amplitudes and thereby the charge current conservation. This
difficulty can be solved in the PRA, where the initial off-shell
gluons and quarks are considered as Reggeons or Reggeized gluons and
quarks, which interact with usual quarks and Yang-Mills gluons  by a
special way, via gauge invariant effective vertices
\cite{Antonov,LipatoVyazovsky}. Our previous study of inclusive jet
\cite{lhcY}, and inclusive prompt photon production
\cite{tevatronY,heraY} show that in the PRA it is enough to consider
only LO $2\to 1$ subprocess to describe inclusive data well.

The LO subprocess, which describe Drell-Yan pair production with
nonzero transverse momentum in the PRA, is annihilation of Reggeized
quark and Reggeized antiquark in lepton pair via virtual photon:
\begin{equation}
{\cal Q}(q_1) +\bar{\cal Q}(q_2)\to \gamma^\star\to l^+(k_1)  +
l^-(k_2)\label{QQDY}.
\end{equation}
Four-momenta of Reggeized quarks (antiquarks) have transverse
components and they read $q_i^\mu=x_iP_i^\mu+q_{iT}^\mu$,
$q_{iT}^\mu=(0,\vec q_{iT},0)$, $q_i^2=q_{iT}^2=-t_i\neq 0$. The
amplitude of the subprocess (\ref{QQDY}) reads in PRA as follows
\begin{equation}
M({\cal Q}_i\bar{\cal Q}_i\to l^+l^-) = 4\pi \alpha e_i \bar
V(x_2P_2)\Gamma^{\gamma,\mu}_{{\cal Q}\bar{\cal
Q}}(q_1,q_2)U(x_1P_1) \otimes \bar U(k_1)\gamma_\mu V(k_2),
\end{equation}
where $e_i$ is the electric charge of quark $i$ (in units of
electron charge), and $\Gamma^{\gamma,\mu}_{{\cal Q}\bar{\cal
Q}}(q_1,q_2)$ is the gauge invariant Fadin-Sherman effective vertex
\cite{FadinSherman,LipatoVyazovsky},
\begin{equation}
\Gamma^{\gamma,\mu}_{{\cal Q}\bar{\cal
Q}}(q_1,q_2)=\gamma^\mu-\frac{2\hat q_1P_1^\mu}{x_2S}-\frac{2\hat
q_2P_2^\mu}{x_1S}.
\end{equation}

We study  the Drell-Yan pair production with nonzero
transverse-momentum in the proton-proton high-energy collisions: $
p(P_1)+ p(P_2) \to l^+(k_1)+ l^-(k_2) + X$, where 4-momenta of
particles are shown in brackets, $l=e,\mu$ (electron or muon),
$q=q_1+q_2=k_1+k_2$  is the 4-momentum of virtual photon,
$Q=\sqrt{q^2}$, $Q_T^2=Q^2+q_T^2=x_1x_2S$. Differential cross
section of this process can be presented as follows:
\begin{eqnarray}
\frac{d\sigma}{dQ^2dq_T^2dyd\Omega}=\frac{\alpha^2}{64\pi^3SQ^4}L_{\mu\nu}W^{\mu\nu},\label{sech0}
\end{eqnarray}
where $y$ is the rapidity of virtual photon (or lepton pair),
$d\Omega=d\phi d\cos\theta$ is the space angle of producing positive
lepton in the rest frame or virtual photon, $\alpha$ is the
electromagnetic constant, $L^{\mu\nu}=2(k_1^\mu k_2^\nu+k_1^\nu
k_2^\mu)-Q^2 g^{\mu\nu}$ is the leptonic tensor, $W_{\mu\nu}=\int
d^4 x e^{iqx}\langle P_1P_2|j_\mu(x)j_\nu(0)|P_1P_2 \rangle$ is the
hadronic tensor, $P_{1,2}=\frac{\sqrt{S}}{2}(1,0,0,\pm 1)$.

The normalized angular distribution of leptons {can be} written
using two different sets of the angular coefficients:
\begin{eqnarray}
&&A_0=\frac{W_L}{W_{TL}}, \quad A_1=\frac{W_\Delta}{W_{TL}}, \quad
A_2=\frac{2W_{\Delta\Delta}}{W_{TL}},
\\
&&\lambda=\frac{2-3A_0}{2+A_0}, \quad \mu=\frac{2A_1}{2+A_0}, \quad
\nu=\frac{2A_2}{2+A_0}.
\end{eqnarray}
Helicity structure functions $W_{T,L,\Delta,\Delta\Delta}$ are
obtained by the projection of hadronic tensor on the photon states
with the different polarizations $\epsilon_\lambda^\mu(q)$,
$\lambda=\pm 1,0$ \cite{Berger2}.

\section{Helicity structure functions in PRA}

We suggest that factorization formula can be used in the PRA:
\begin{eqnarray}
d\sigma (pp\to l^+l^-X)=\sum_{q}\int \frac{d\phi_1}{2\pi} dt_1
\frac{dx_1}{x_1} \frac{d\phi_2}{2\pi} dt_2 \frac{dx_2}{x_2}
 \Phi^p_q(x_1,t_1,\mu)\Phi^p_{\bar q}(x_2,t_2,\mu)
d\hat\sigma({\cal Q}\bar {\cal Q}\to l^+l^-)\label{DYkt},
\end{eqnarray}
where $\Phi^p_q(x_{1,2},t_{1,2},\mu)$ {are} unitegrated over the
transverse momentum PDFs of the Reggeized quarks. In our numerical
analysis, we adopt the prescription proposed by Kimber, Martin, and
Ryskin (KMR) \cite{KMR} to obtain unintegrated quark PDFs of the
proton from the conventional integrated one. The differential cross
section $\hat\sigma({\cal Q}\bar {\cal Q}\to l^+l^-)$ {is} directly
connected with squared matrix element. In the PRA we obtain for the
squared amplitude of the subprocess (\ref{QQDY})
\begin{equation}
\overline{|M({\cal Q}_i\bar {\cal Q}_i\to
l^+l^-)|^2}=\frac{16\pi^2}{3Q^4}\alpha^2e_i^2
L^{\mu\nu}w_{\mu\nu}^{Regge}\label{qtensor},
\end{equation}
where the {partonic} tensor {for} Reggeized quarks reads:
\begin{eqnarray}
&w_{\mu\nu}^{Regge}=x_1x_2\bigl[-Sg^{\mu\nu}+2(P_1^\mu P_2^\nu+
P_2^\mu
P_1^\nu)\frac{(2x_1x_2S-Q^2-t_1-t_2)}{x_1x_2S}+\\
&+\frac{2}{x_2}(q_1^\mu P_1^\nu+q_1^\nu
P_1^\mu)+\frac{2}{x_1}(q_2^\mu P_2^\nu+q_2^\nu P_2^\mu)+
\frac{4(t_1-x_1x_2S)}{Sx_2^2}P_1^\mu P_1^\nu+
\frac{4(t_2-x_1x_2S)}{Sx_1^2}P_2^\mu
P_2^\nu\bigr]\label{Qtensor}.\nonumber
\end{eqnarray}
The helicity structure functions $W_{T,...}^{Regge}$ at the fixed
values of variables  $S,Q,q_T,y$ can be presented via corresponding
quark helicity functions $w^{Regge}_{T,...}$:
\begin{equation}
W_{T,...}^{Regge}(S,Q,q_T,y)=\frac{8\pi^2S}{3Q_T^4}\int dt_1\int
d\phi_1 \sum_q \Phi_q^p(x_1,t_1,\mu^2)\Phi_{\bar q}^p(x_2,t_2,\mu^2)
w^{Regge}_{T,...},
\end{equation}
where
\begin{eqnarray}&&w_T^{Regge}=Q^2+\frac{(\vec q_{1T}+\vec q_{2T})^2}{2},
\quad w_L^{Regge}=(\vec q_{1T}-\vec q_{2T})^2,\nonumber \\
&&w_\Delta^{Regge}=0, \quad w_{\Delta\Delta}^{Regge}=\frac{(\vec
q_{1T}+\vec q_{2T})^2}{2}.\end{eqnarray} {Factorization scale is
chosen to be $\mu=\xi Q_T$, and $\xi$ was varied between $1/2$ and
2, to obtain the scale uncertainty. It was found, that observables
$A_0,\ A_2,\ \lambda,\ \nu$ are much more stable under the scale
variations than values of cross sections. Their variations under
scale changing in the kinematical regions under consideration are
found to be less than 20 \%.}

 Our LO PRA results should be corrected by the so-called
K-factor, which includes higher order (HO) QCD corrections to the LO
diagrams. The main part of HO corrections arising from real gluon
emission already accounted in LO PRA. Another part comes from the
non-logarithmic loop corrections arising from gluon vertex
corrections. Accordingly to Ref.~\cite{Kfactor}, this K-factor is
written as follows $ K({\cal Q}\bar{\cal Q}\to
\gamma^*)=\exp(C_F\frac{\alpha_s(\mu^2)}{2\pi}\pi^2)$, where a
particular scale choice $\mu^2=Q_T^{4/3}Q^{2/3}$ has been used to
evaluate $\alpha_s(\mu^2)$.

\section{Results}
\begin{figure}
\begin{tabular}{cc}
\includegraphics[scale=0.35]{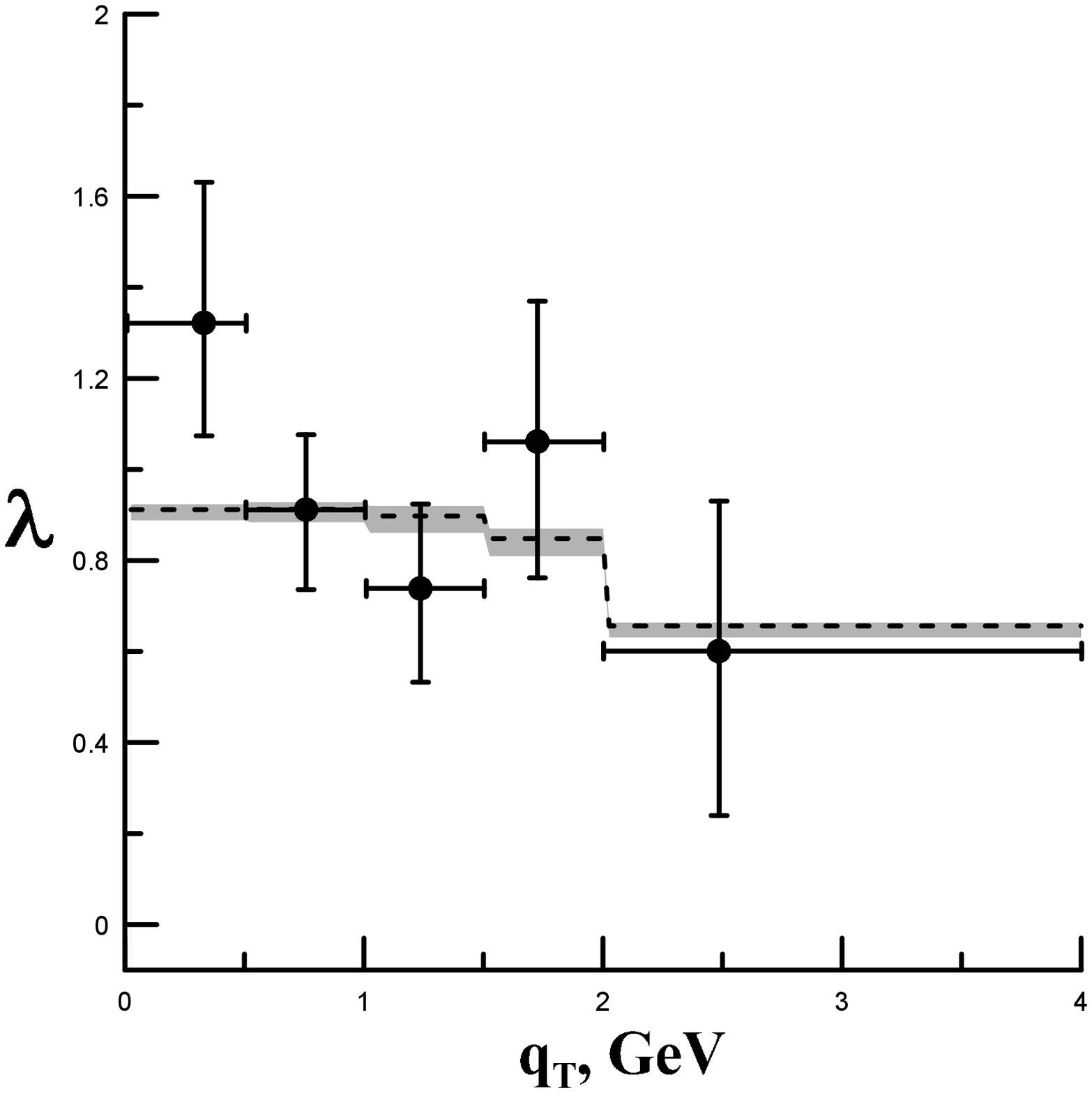} & \includegraphics[scale=0.35]{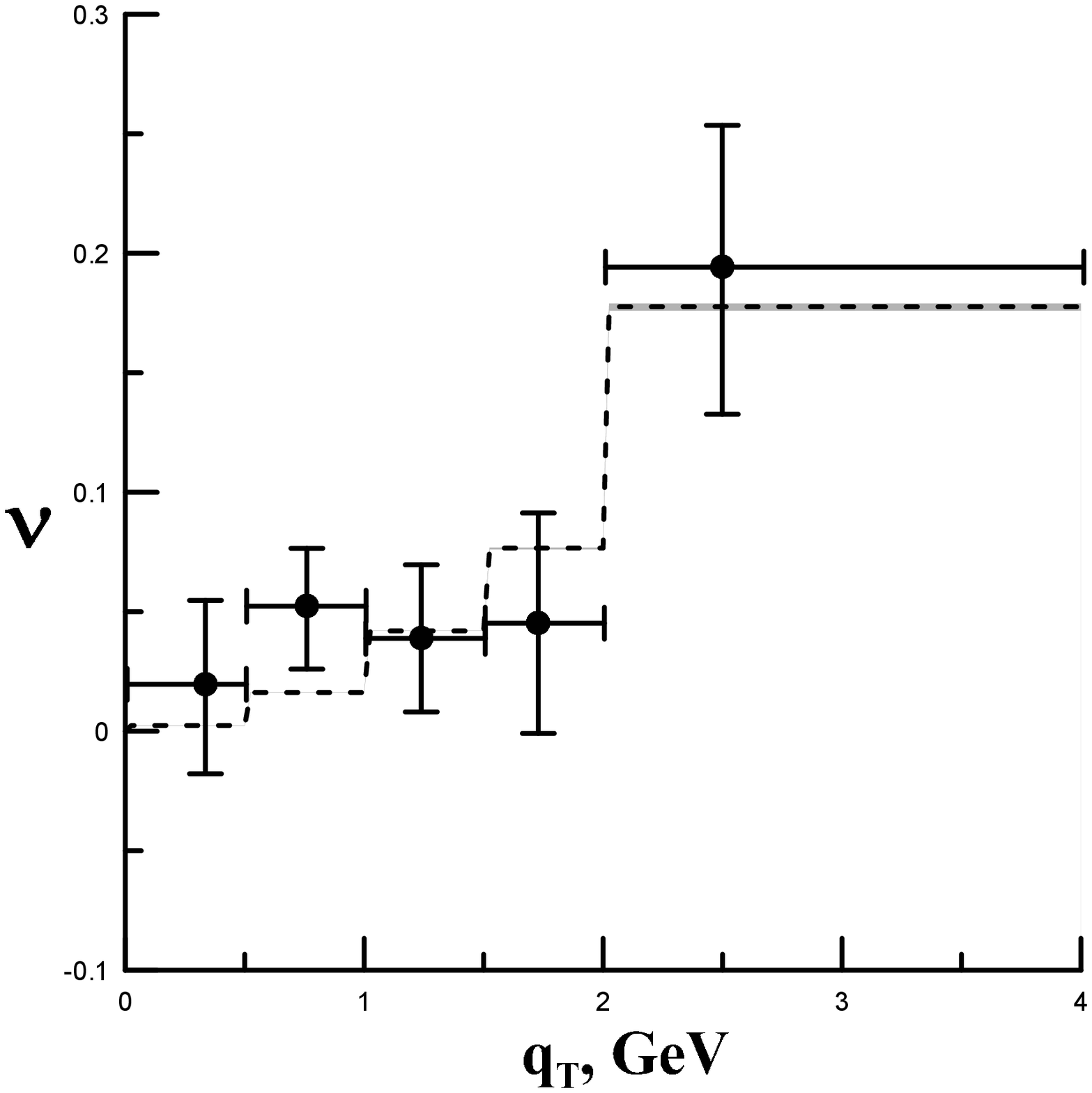}\\
\end{tabular}
\caption{Angular coefficients $\lambda$ (left panel) and $\nu$
(right panel)  as a function of $q_T$. The histogram corresponds to
LO calculation in the PRA with KMR \cite{KMR} unintegrated PDFs. The
data are from NuSea Collaboration \cite{NuSea}.} \label{fig1}
\end{figure}

Recently, NuSea Collaboration from Fermilab Tevatron has published
data  for Drell-Yan pair production in fixed-target experiment with
hydrogen and deuterium targets at $E_p=800$ GeV proton beam
($\sqrt{S}=39$ GeV) \cite{NuSea}. The measurements were made in the
following kinematics domain: $4.5<Q<15$ GeV, $0<q_T<4$ GeV,
$0<x_F<0.8$. The results of measurements of angular distributions
are presented in terms of angular coefficients  $\lambda$ and  $\nu$
as functions of virtual photon transverse momentum. We find good
agreement of our LO PRA calculations with data for $\lambda$ and
$\nu$ at all values of $q_T$, as it is shown in Fig.~\ref{fig1}.
Additionally, we  predict $\mu=0$ that is also in agreement with
data with the experimental accuracy.
\section*{Acknowledgements} M.~N. thanks Organizing Committee of HSQCD
2012 for invitation and nice accommodation during the Conference.
The work was supported by the Ministry for Science and Education of
the Russian Federation under Contract No.~14.740.11.0894. The work
of M.~N. is supported also by the Grant of the Student's Stipend
Program of the Dynasty Foundation.

\end{document}